\documentclass[english,prd,superscriptaddress,
               preprintnumbers,twocolumn,noshowpacs]{revtex4-1}
\usepackage[latin1]{inputenc}
\usepackage{amsmath}
\usepackage{amssymb}
\usepackage{graphicx}
\usepackage[caption=false]{subfig}
\usepackage{amsthm}
\usepackage{bm}
\usepackage{color}
\makeatletter
\@ifundefined{textcolor}{}
{%
 \definecolor{BLACK}{gray}{0}
 \definecolor{WHITE}{gray}{1}
 \definecolor{RED}{rgb}{1,0,0}
 \definecolor{GREEN}{rgb}{0,1,0}
 \definecolor{BLUE}{rgb}{0,0,1}
 \definecolor{CYAN}{cmyk}{1,0,0,0}
 \definecolor{MAGENTA}{cmyk}{0,1,0,0}
 \definecolor{YELLOW}{cmyk}{0,0,1,0}
 }
\usepackage{amsfonts}\usepackage{dcolumn}
\usepackage{babel}

\newcommand{\fnl}{f_{\mathrm{NL}}}

\newcommand{\gnl}{g_{\mathrm{NL}}}
\newcommand{\be}{\begin{equation}}
\newcommand{\ee}{\end{equation}}
\def\bs{\begin{subequations}}
\def\es{\end{subequations}}
\newcommand{\vect}[1]{\bm{\mathrm{{#1}}}}

\usepackage[T1]{fontenc} 

\usepackage{braket}

\newcommand\ben{\begin{enumerate}}
\newcommand\een{\end{enumerate}}
\newcommand\bal{\begin{align*}}
\newcommand\eal{\end{align*}}
\newcommand\bi{\begin{itemize}}
\newcommand\ei{\end{itemize}}

\def\id{\protect{{1 \kern-.28em {\rm l}}}}

\newcommand{\gae}{\lower 2pt \hbox{$\, \buildrel {\scriptstyle >}\over {\scriptstyle
\sim}\,$}}
\newcommand{\lae}{\lower 2pt \hbox{$\, \buildrel {\scriptstyle <}\over {\scriptstyle
\sim}\,$}}

\begin{document}

\title{Generating the cosmic microwave background power asymmetry with $g_{NL}$}

\author{Zachary Kenton}
\email{z.a.kenton@qmul.ac.uk}
\author{David J. Mulryne}
\email{d.mulryne@qmul.ac.uk}
\author{Steven Thomas}
\email{s.thomas@qmul.ac.uk}

\affiliation{School of Physics and Astronomy, Queen Mary University of London,\\
Mile End Road, London, E1 4NS,  UK.}

\date{\today}

\begin{abstract}
We consider a higher order term in the $\delta N$ expansion for the CMB power asymmetry generated by a superhorizon isocurvature field fluctuation. The term can generate the asymmetry without requiring a large value of $\fnl$. Instead it produces a non-zero value of $\gnl$. A combination of constraints leads to an allowed region in $\fnl-\gnl$ space. To produce the asymmetry with this term without a large value of $\fnl$ we find that the isocurvature field needs to contribute less than the inflaton towards the power spectrum of the curvature perturbation.
\end{abstract}

\maketitle

\section{Introduction}

Inflation is widely accepted as the likely origin for structure in our universe.
Its generic predictions of a nearly scale invariant and close to Gaussian primordial 
curvature perturbation, $\zeta$, have been confirmed with increasing precision 
by successive Cosmic Microwave Background (CMB) experiments. There are, however, 
also observational anomalies which are harder to explain within the standard inflationary paradigm. 
One such anomaly is the hemispherical power asymmetry -- the observation that for scales with $l \leq 60 $ there is more power in CMB temperature fluctuations in one half of the sky than the other. First identified in the Wilkinson Microwave Anisotropy Probe data \cite{Eriksen2003,Eriksen2007,Hansen2004,Hoftuft2009}, it 
was later confirmed by the Planck collaboration \cite{PlanckCollaboration2013} and others \cite{Paci2013,Flender2013,Akrami2014}, although its significance remains disputed \cite{2011ApJS..192...17B}. In this work we treat the asymmetry as a real effect which requires a primordial origin. So far, CMB data has been fitted to a template which models the asymmetry as a spatially linear modulation. 

The leading primordial explanation for this asymmetry is the Erickcek-Kamionkowski-Carrol (EKC) mechanism \cite{Erickcek2008,Erickcek2008a}, in which a long-wavelength isocurvature perturbation modulates the power on shorter scales. Further work investigating this effect includes Refs.~\cite{Kobayashi2015,Liddle2013,Dai2013,Lyth2013,Namjoo2013,Namjoo2014,Lyth2014a,Kanno2013,Abolhasani2013}.
The origin of the long wavelength mode may be explicitly realised in the open inflation scenario of \cite{Liddle2013} or due to a domain wall, as in, for example, \cite{Jazayeri:2014nya,Kohri:2013kqa}. 

The $\delta N$ formalism provides a convenient expression for  
the modulation of power by a super-horizon mode, as reviewed below. In principle many terms in this $\delta N$ expansion can contribute to the observed asymmetry. Until now, however, most theoretical work has focused on the leading term, which can have the form of a spatially linear modulation. 

If the leading term in $\delta N$ is responsible for the asymmetry then a further consequence is that the local bispectrum parameter must satisfy the constraint $\fnl\gae 30\beta$ \cite{Kobayashi2015} \footnote{Without considering our position within the modulation, and with slightly different numerical values \cite{Kanno2013} earlier found $\fnl\gae 66$.} on the scales that are modulated. A value of $\beta<1$ can be achieved but only if our observable universe is located at a fine-tuned region within the long-wavelength perturbation \cite{Kobayashi2015}, and otherwise can be much larger than one. Combined temperature and polarization data bounds a purely scale-independent local bispectrum as $\fnl = 0.8\pm 5.0$ at $68\% \text{CL}$ \cite{Ade:2015ava}, while we work with $\fnl \lae 10$ as a rough $95\% \text{CL}$. The asymmetry appears to be scale dependent \cite{Hirata2009}, and hence the non-Gaussianity produced must also be, but there are no direct constraints on such  a strongly scale dependent non-Gaussianity. A new parametrisation of the scale-dependence of the non-Gaussianity and its application to the scale-dependence of the asymmetry was given in \cite{Byrnes:2015asa}, which includes an accompanying $\gnl$. It is, however, perhaps unlikely that a very large value of $\fnl$ could be accommodated by current observations, even if $\fnl$ decays with scale. 

In this short paper, therefore, we investigate whether the next term in the $\delta N$ expression for the asymmetry could instead be responsible. 
We find it can, without violating any other observational or 
self-consistency constraints. It contributes a more general modulation of the power, leading to an asymmetry, which does not necessarily only involve a spatially linear modulation \footnote{To the best of our knowledge, no current data analysis has been performed using a template involving these more general modulation terms.}.  Using this higher order term requires a non-zero value of $\gnl$, but allows for a smaller value of $f_{NL}$ than when the linear term alone contributes. If this higher order term is responsible for the asymmetry, then the allowed parameter space indicates the modulating isocurvature field must contribute less than the inflaton towards the total power spectrum of the curvature perturbation on scales which are modulated, and this may be considered a fine-tuning of the model. Related to this, we find that if this higher order term is dominant in our observable patch, then in certain neighbouring patches the linear term will instead be dominant. 

In this paper, as a first step we only focus on one of the higher order terms, but the idea is more general and could be applied to a combination of higher order terms. Satisfying the constraints in that case might be more complicated than the simple use of exclusion plots that we employ here.

\section{Generating the Asymmetry}
\subsection{The $\delta N$ Formalism}  

Our calculation is performed within the $\delta N$ formalism \cite{Sasaki1995,Sasaki1998,Wands2000,Lyth2004,Lyth2005} 
 which states that $\zeta$ can be associated with the 
difference in the number of e-folds undergone by neighbouring positions in the 
universe from 
an initial flat hypersurface at horizon crossing to a final uniform density one 
when the dynamics have become adiabatic: $ \zeta = \delta N$. 
On the flat hypersurface 
the inflationary fields are not constant, and by writing $N$ as a 
function of the fields, $\delta N$ can be written as a 
Taylor expansion in the horizon crossing field fluctuations.

We consider two scalar fields, though our work easily generalises for more than two fields, and we take both our fields to have canonical kinetic terms. 
We choose the inflaton field, denoted $\phi$, to be the direction 
in field space aligned with the inflationary trajectory at horizon exit, 
$t_{*}$, so that $\epsilon^*=\epsilon_\phi^*$ and this implies the derivative 
of $N$ with respect to the inflaton is a constant
\begin{align}
N_{,\phi} = (2\epsilon^*)^{-1/2}. \label{nphi}
\end{align}
The isocurvature field orthogonal to $\phi$ is denoted $\chi$, and the curvature perturbation has contributions from both fields
\begin{align}
 \zeta = N_{,\phi}\delta{\phi} + N_{,\chi}\delta{\chi}  + \frac{1}{2}N_{,\chi\chi}\delta{\chi}^2 + \frac{1}{6}N_{,\chi\chi\chi}\delta{\chi}^3 +... \label{eq:deltaNzetachis}
 \end{align} 
where we have neglected terms with higher order $\phi$ derivatives since they
 are negligible. The arguments of $N$ and its derivatives are usually 
 taken to be the average values of the fields within our observable universe, 
 denoted $\phi_0$ and $\chi_0$, while $\delta \phi$ and $\delta \chi$ contain all 
 fluctuations in $\phi$ and $\chi$ with wavelengths of order the size of our observable 
 universe or less.

The power spectrum of the curvature perturbation is then given by
\begin{align}
P_\zeta &= N_{,I}N_{,I}\left ( \frac{H}{2\pi} \right )^2 \label{pzetaapprox}
\end{align}
where $I$ runs over $\{\phi, \chi\}$, the summation convention has been used, and we have neglected higher order $\delta\phi$ and $\delta\chi$ correlators.

Non-Gaussianities in $\zeta$ are generated because of the non-linear relationship between $\zeta$ and $\delta\chi$ in \eqref{eq:deltaNzetachis}. In particular, one finds for the local bispectrum, $f_{NL}$, and trispectrum, $g_{NL}$, parameters that \cite{Byrnes:2006vq,Seery:2006js}
\begin{align}
f_{NL} = \frac{5}{6}\frac{N_{,\chi\chi}N_{,\chi}^2}{(N_{,I}N_{,I})^2} \label{fnl}
\\
g_{NL} = \frac{25}{54}\frac{N_{,\chi\chi\chi}N_{,\chi}^3}{(N_{,I}N_{,I})^3} \label{gnl}\,.
\end{align}
In what follows we will only be concerned with the magnitude of $f_{NL}$ and $g_{NL}$ , $|f_{NL}|$ and $|g_{NL}|$, but to avoid clutter we will drop the absolute symbols. 
We will also use the expression for the tensor-to-scalar ratio
\begin{align}
r = \frac{8}{N_{,I}N_{,I}} \label{rnini}
\end{align}
and we will find it convenient to define the contribution of $\chi$ to the total power spectrum
\begin{align}
x\equiv \frac{P_\chi}{P_{\zeta}} = \dfrac{N_{,\chi}^2}{N_{,I}N_{,I}} = 1 - \frac{r}{16\epsilon^*}. \label{xdef}
\end{align}

\subsection{ Superhorizon Fluctuation}
In addition to the background value of the fields inside our observable universe, $\{\phi_0, \chi_0\}$ and their fluctuations with 
wavelength inside our observable universe, $\{ \delta \phi, \delta \chi \}$, the EKC mechanism works by postulating a superhorizon field fluctuation in $\chi$, denoted $\Delta\chi(\vect{x})$, with wavelength, $k_L^{-1}$, much larger than the size of our observable universe, this size given by the distance to the last scattering surface, $x_{d}$, such that $k_Lx_{d} \ll 1$. 
We assume the leading order behaviour $\Delta\chi(\vect{x})=\overline{\Delta\chi}(\vect{\hat{n}}\cdot \vect{\hat{k}_L})$ for $\vect{x}$ within our observable universe, where $\vect{\hat{n}} = \vect{x}/|\vect{x}|$ and $\vect{\hat{k}_L} = \vect{k_L}/|\vect{k_L}|$,
and we don't assume any particular form for the 
fluctuation outside of our observable patch. Note that in this paper we take $\overline{\Delta\chi}$ to be the maximum variation in $\chi$
across our patch about our observable universe's average field value $\chi_0$ as seen in the left panel of Fig~\ref{fig1} \footnote{
These properties are in contrast to the $\Delta\sigma(\vect{x})$ of Ref.~\cite{Kobayashi2015} which is a long wavelength fluctuation around the background field value, $\sigma_{bg}$, of the entire universe which is much larger than our observable patch. 
Our results can be matched to the results of their section 6, with our $\chi_0$ replacing their $\sigma_{bg} + \Delta\tilde{\sigma}\cos\theta$, and our $\overline{\Delta\chi}$ replacing their $k_Lx_d\Delta\tilde{\sigma}\sin\theta$. One might demand $|N_{,\chi\chi\chi}\left (\Delta\tilde{\sigma} \right )^3|<1$ which is in fact satisfied by \eqref{octnxxx} for $k_L x_d\approx 0.1$.
}.

Superhorizon fluctuations source multipole moments in the CMB, upon which there are constraints from the observed homogeneity of the universe \cite{Erickcek2008a,Erickcek2008}. Using the non-linear results of \cite{Kobayashi2015},  together with the multipole constraints from \cite{Erickcek2008a}, we have the following homogeneity constraints from the quadrupole and octupole respectively \footnote{The hexadecapole will also receive contributions from $N_{,\chi\chi\chi}$ though it appears suppressed by powers of $k_L x_d$ meaning if it satisfies the octupole constraint it will also satisfy the hexadecapole, and similarly for higher derivatives and multipoles.}
\begin{align}
|N_{,\chi\chi}\left ({{\overline{\Delta\chi}}}\right )^2 | &< 1.1  \times10^{-4} \label{quadnxx}
\\
|N_{,\chi\chi\chi}\left ({{\overline{\Delta\chi}}}\right )^3| &< 8.6\times10^{-4} \label{octnxxx}
\end{align}
where we have assumed no cancellation between $\delta N$ terms.
We also take the following constraint
\begin{align}
|N_{,\chi}{{\overline{\Delta\chi}}}| &< a P_{\zeta}^{1/2} \label{dipoleanyway}
 \end{align}
 where $P_{\zeta}=2.2\times10^{-9}$ \cite{Ade:2013zuv} and $a$ is some threshold parameter.
 
\subsection{Asymmetry}

The superhorizon fluctuation modulates the power spectrum on shorter scales, and so it depends on the direction $\vect{\hat{n}}$ through
\begin{align}
P_\zeta[\vect{\hat{n}}] &= P_\zeta[\chi_0 + \Delta\chi(\vect{\hat{n}})]. \label{Pani}
\end{align}
Since $\Delta\chi(\vect{\hat{n}})<\overline{\Delta\chi}$ in our patch, and $\overline{\Delta\chi}$ is small, we can Taylor expand $P_{\zeta}$ in \eqref{Pani} in powers of $\Delta\chi(\vect{\hat{n}})$ giving
\begin{align}
P_\zeta[\vect{\hat{n}}] = P_\zeta\left (1 + 2\sum_{m=1}^{\infty}A_m (\vect{\hat{n}}\cdot \vect{\hat{k}_L})^m \right ) \label{Ptaylorani}
\end{align}
where the round brackets indicate multiplication, 
\begin{align}
A_m \equiv \frac{1}{2P_\zeta} \frac{ (\overline{\Delta\chi})^m}{m!} \frac{\partial ^m P_\zeta}{\partial \chi ^m} \label{Adef}
\end{align}
and we have used the shorthand that when $P_\zeta$ and its derivatives appear without an argument they are taken to be evaluated at the average field values of the observable universe. Observations indicate a power asymmetry, with the power along the preferred direction $\vect{\hat{n}}=\vect{\hat{k}_L}$ being greater than the power on the opposite side of the sky $\vect{\hat{n}}=-\vect{\hat{k}_L}$. We note that only the odd $m$ terms in \eqref{Ptaylorani} can contribute towards an asymmetry of this sort, with the even terms contributing only towards general anisotropy. 

Usually only the $m=1$ term is kept, and the data has been fitted to this with the result that \cite{PlanckCollaboration2013} $A_1=0.07$. The $m=1$ and $m=2$ terms were considered in \cite{Byrnes:2015asa} \footnote{Although the authors of \cite{Byrnes:2015asa} claim the limit $2A_2=0.002\pm0.016$, which they inferred from \cite{Kim:2013gka}, we think \cite{Kim:2013gka} only constrains a term in Fourier rather than real space $(\vect{\hat{k}}\cdot \vect{\hat{E}}_{\text{cl}})^2$, and so to the best of our knowledge there is no direct bound on $A_2$. }. Here we consider instead the $m=3$ term, since this can contribute towards asymmetry \footnote{One might worry that the second and third order terms in \eqref{eq:deltaNzetachis} become of comparable size for asymmetry generated by the $m=3$ term and so loop corrections to $\fnl$ may be important, changing the expression for $\fnl$ in \eqref{fnl}. However one can check these loop corrections to $\fnl$ are subdominant to the tree level term for $\overline{\Delta\chi}>\delta\chi$.  }. Ideally a fit to the data with $m=1,2,3$ terms should be done to constrain the parameters $A_1, A_2$ and $A_3$. In the absence of this, we will look at the simplest case involving only the $m=3$ term and take \footnote{We expect this to be $\gae 0.07$ since the area under a cubic curve is less than the area under a linear curve if they share the same endpoints.} $A_3 \gae 0.07$.

\subsection{Linear Term Asymmetry}
It has been noted in e.g \cite{Kanno2013,Lyth2013,Kobayashi2015} that a large $f_{NL}$ accompanies the asymmetry when only the $m=1$ term is considered, and we briefly review this now. Differentiating \eqref{pzetaapprox} gives
\begin{align}
A_1= \frac{N_{,\chi\chi}N_{,\chi}\overline{\Delta\chi}}{N_{,I}N_{,I}}.
\end{align}
We now combine this with constraint \eqref{quadnxx} giving
\begin{align}
f_{NL} \approx \frac{5N_{,\chi\chi}N_{,\chi}^2}{6(N_{,I}N_{,I})^2} \gae 37\left ( \dfrac{A_1}{0.07}\right )^2, \label{fnlbadekc}
\end{align}
which is outside of the observational bounds for a local-type non-Gaussianity \footnote{Different authors have used slightly different values for the quadrupole and octupole, and the value of $A$, leading to other numbers appearing in \eqref{fnlbadekc} ranging from $30-70$.}.

\subsection{Cubic Term Asymmetry}
The asymmetry may be due to multiple odd $m$ terms in \eqref{Ptaylorani}. We will now show that postulating the cubic $m=3$ term is dominant over the linear $m=1$ term, and is responsible for the asymmetry, allows the constraint on $f_{NL}$ to be relaxed, but introduces new ones on $g_{NL}$. Later we will check the self-consistency of ignoring the $m=1$ term compared to the $m=3$ one.

Differentiating \eqref{pzetaapprox} three times gives
\begin{align}
\frac{{P_\zeta}_{,\chi\chi\chi}}{{P_\zeta}} = \frac{6N_{,\chi\chi\chi}N_{,\chi\chi} +  2N_{,\chi\chi\chi\chi}N_{,\chi}}{N_{,I}N_{,I}}.\label{pzeta3}
\end{align}
We will be interested in the case where the asymmetry is generated by the $N_{,\chi\chi\chi}N_{,\chi\chi}$ term, and we neglect $N_{,\chi\chi\chi\chi}$, so that our asymmetry is given by \footnote{We consider the $N_{,\chi\chi\chi\chi}N_{,\chi}$ term in the conclusion, noting that this term may avoid a large $\fnl$, introducing a non-zero $h_{\text{NL}}$ instead.}
\begin{align}
A_3 = \frac{N_{,\chi\chi\chi}N_{,\chi\chi}(\overline{\Delta\chi})^3 }{2N_{,I}N_{,I}}. \label{asymmetron}
\end{align}
In this case, we now show there is still a lower bound on $f_{NL}$, but this time it depends on $x$ defined in \eqref{xdef}.  Using \eqref{asymmetron} together with the octupole constraint \eqref{octnxxx}, we find 
\begin{align}
f_{NL} \approx \frac{5N_{,\chi\chi}N_{,\chi}^2}{6(N_{,I}N_{,I})^2} \gae 9.5	\left ( \dfrac{A_3}{0.07}\right )\left ( \dfrac{x}{0.07}\right ). \label{fnlbadatron}
\end{align}
We see that if $x$ is sufficiently small, we can have an acceptably small $f_{NL}$ in this scenario. We will later show that there is a lower bound $x\gae A_3$, and so $9.5$ is the smallest value of $f_{NL}$ allowed from this cubic term alone \footnote{Although one can get a value of $7$ if both $m=1,3$ terms contribute equally $A_1=A_3=0.035$.}, which is an improvement compared to the contribution from the linear term alone.

\subsection{Consistency Checks}
For simplicity we assumed that the asymmetry is only due to the $m=3$ term in \eqref{Ptaylorani}, which then must be larger than the $m=1$ term. We therefore require 
\begin{align}
\frac{N_{,\chi\chi\chi}N_{,\chi\chi}(\overline{\Delta\chi})^2}{2N_{,\chi\chi}N_{,\chi}} >1.\label{cubicvvslin}
\end{align}
Even powers of $\overline{\Delta\chi}$ don't contribute towards the asymmetry 
but they do still cause more general anisotropy of the power spectrum.
Since these anisotropies have not been observed, we also demand the following
\begin{align}
\frac{N_{,\chi\chi\chi}N_{,\chi\chi}\overline{\Delta\chi}}{N_{,\chi\chi\chi}N_{,\chi}} 
 >b\label{cubicvsquad1}
\\
\text{and }\frac{N_{,\chi\chi\chi}N_{,\chi\chi}\overline{\Delta\chi}}{N_{,\chi\chi}^2} 
 >c \label{cubicvsquad2}
\end{align}
where $b,c$ are some threshold parameters.

There is a lower limit on $x=x(\chi_0)$ coming from $x(\chi_0-\overline{\Delta\chi})>0$, by definition \eqref{xdef}. Expanding out $x(\chi_0-\overline{\Delta\chi})$ to cubic order and neglecting the linear term, we find $x(\chi_0) \gae A_3$ for $b,c \sim \mathcal{O}(1)$.

\subsection{Allowed Parameters}
We have six constraints to simultaneously satisfy: \eqref{quadnxx}, \eqref{octnxxx}, \eqref{dipoleanyway}, \eqref{cubicvvslin}, \eqref{cubicvsquad1} and \eqref{cubicvsquad2}. Using \eqref{asymmetron} to substitute for $\overline{\Delta\chi}$, and using \eqref{fnl}, \eqref{gnl} and \eqref{rnini} the six constraints become, respectively,
\begin{align}
g_{NL} &> \left ( \dfrac{x}{0.07}\right ) \left ( \dfrac{A_3}{0.07}\right )4.3\times10^{3}f_{NL}^{1/2} \label{gf1}
\\
f_{NL} &>\left ( \dfrac{x}{0.07}\right ) \left ( \dfrac{A_3}{0.07}\right )9.5\label{gf2}
\\
g_{NL} &>  \left ( \dfrac{x}{0.07}\right )^4\left ( \dfrac{A_3}{0.07}\right ) 1.8\times10^{7} a^{-3/2} f_{NL}^{-1}\label{gf3}
\\
g_{NL} &> \left ( \dfrac{x}{0.07}\right ) \left ( \dfrac{A_3}{0.07}\right )^{-2}19 f_{NL}^{2}\label{gf4}
\\
g_{NL} &<\left ( \dfrac{x}{0.07}\right )^{-2}\left ( \dfrac{A_3}{0.07}\right ) 19 b^{-3}f_{NL}^{2}\label{gf5}
\\
g_{NL} &>\left ( \dfrac{x}{0.07}\right )^{-1/2}\left ( \dfrac{A_3}{0.07}\right )^{-1/2}  6.7 c^{3/2}f_{NL}^{2}\label{gf6}.
\end{align}
\begin{figure}
\centering
{\mbox{\includegraphics[width=0.24\textwidth]{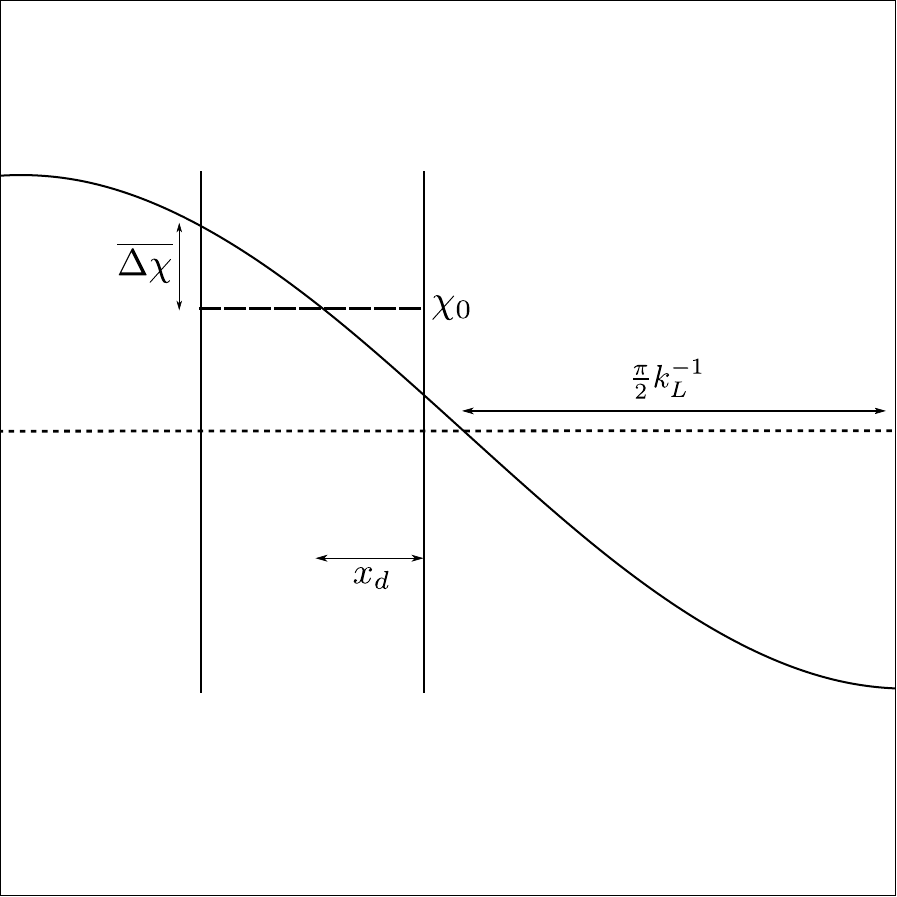}
\includegraphics[width=0.24\textwidth]{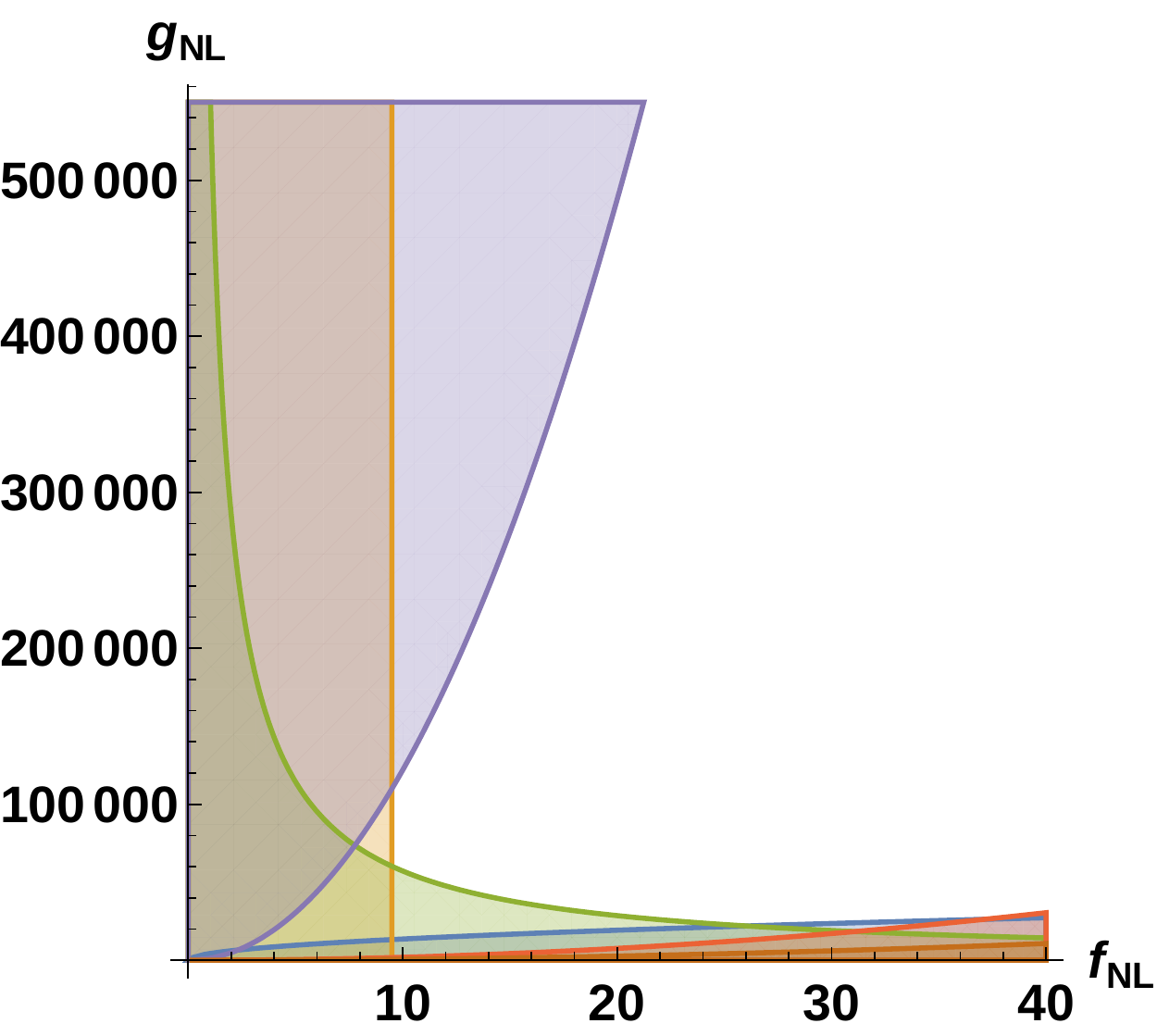} } }
   \caption{\textit{Left}: The value of $\chi$ varies by an amount $\overline{\Delta\chi}$ from its average value, $\chi_0$, within our observable universe (the interior of the two vertical lines), due to the long-wavelength fluctuation (solid wave). The average within our observable universe (long-dashed line) is not necessarily the same as the background value over the entire universe (dotted horizontal line). \textit{Right}: Exclusion plot for \eqref{gf1}-dark-blue, \eqref{gf2}-orange, \eqref{gf3}-green, \eqref{gf4}-red, \eqref{gf5}-purple, \eqref{gf6}-brown, with $x=0.07$, $a=10$, $b=0.25$ and $c=1$. The red and brown lines are hard to see on this scale at the bottom of the plot. The allowed region is left white.}
\label{fig1}
\end{figure}
In the right panel of Fig~\ref{fig1} we plot the allowed region, left in white, for \eqref{gf1}-\eqref{gf6}, with $x=A_3=0.07$, $a=10$, $b=0.25$ and $c=1$. We find that the cubic term can generate the required asymmetry with a lower value of $f_{NL}$ than from the linear term alone. Moreover it requires a non-zero value of $g_{NL} \gae 5\times 10^4$ for the smallest allowed values of $\fnl$. Note that if $x$ is much bigger than $0.07$ then this pushes the allowed values of $\fnl$ and $\gnl$ up. The small value of $x=0.07$ may be considered a fine-tuning required when only the $m=3$ term generates the asymmetry.

\subsection{Outside Our Observable Patch}
In the above we neglected the first order $m=1$ term in \eqref{Ptaylorani}, assuming that this term is small in our observable universe. However, since we are considering a scenario in which $N_{,\chi\chi}$ and $N_{,\chi\chi\chi}$ are non-zero,  neighbouring regions of the universe with a different background field value
may have a larger first order term. This is closely related to a similar effect in inhomogeneous non-Gaussianity \cite{2012JCAP...03..012B,2013PhRvL.110m1301N,Nurmi:2013xv,LoVerde:2013xka,Baytas:2015nja}. If this term is larger in neighbouring patches this would not violate observational bounds, but would imply that our position within neighbouring regions was finely tuned -- in the sense that neighbouring regions would instead see a dominant first order term. Although not invalidating the proposed scenario, it would make it less appealing.
The biggest change in the average value of $\chi$ is in a neighbouring patch along the direction of the long wavelength mode, where its average value is of order $\chi_0+\overline{\Delta\chi}$, since $\overline{\Delta\chi} > \delta\chi$. The first order term in these patches is then of order
\begin{align}
\begin{split}
{N_{,\chi\chi}N_{,\chi}}\Big|_{\chi_0 + \overline{\Delta\chi}} = &{N_{,\chi\chi}N_{,\chi}}\Big|_{\chi_0} 
\\
&+ \overline{\Delta\chi}(N_{,\chi\chi}^2+ N_{,\chi\chi\chi}N_{,\chi} )\Big|_{\chi_0}
\\
&+ \dfrac{3}{2}(\overline{\Delta\chi})^2{N_{,\chi\chi\chi}N_{,\chi\chi}}\Big|_{\chi_0}+...
\end{split} \label{nccncpatch}
\end{align}
where we have neglected fourth and higher derivatives of $N$. The order $\overline{\Delta\chi}$ term in \eqref{nccncpatch} is related to the zeroth order term by
\begin{align}
\frac{\overline{\Delta\chi}{(N_{,\chi\chi}^2+ N_{,\chi\chi\chi}N_{,\chi} )}}{N_{,\chi\chi}N_{,\chi}}  \Big|_{\chi_0} > \left (b+c \right ) 
\end{align}
and so these terms are of comparable order for $b,c=\mathcal{O}(1)$ and if \eqref{cubicvsquad1} and \eqref{cubicvsquad2} are not hierarchical inequalities.
The order $(\overline{\Delta\chi})^2$ term in \eqref{nccncpatch} is related to the zeroth order term using \eqref{cubicvvslin}
\begin{align}
\frac{3(\overline{\Delta\chi})^2{N_{,\chi\chi\chi}N_{,\chi\chi}}}{2N_{,\chi\chi}N_{,\chi}}  \Big|_{\chi_0} >3
\end{align}
so we see that the first order term in $\overline{\Delta\chi}$ in \eqref{Ptaylorani} in these neighbouring patches will actually be of the same order or larger than the cubic one in our own patch which we consider to be repsonsible for the asymmetry.
This then implies that in these neighbouring patches the value of $f_{NL}$ is necessarily larger than in our own patch. This agrees with the result of \cite{Byrnes:2013qjy} that if $g_{NL} \gg f_{NL}$ in our observable patch, then neighbouring patches will generically have a larger value of $f_{NL}$ than in our own. If the asymmetry in our patch is due to the third order term rather than the linear term, then our patch should be considered fine-tuned compared to its neighbours along the direction of the long wavelength mode.

\section{Conclusion}

We have presented a mechanism involving a modulating isocurvature field which can produce 
the required hemispherical power asymmetry while satisfying the homogeneity constraints, and which produces non-Gaussianity within observational bounds. 
A novel feature is the non-zero value of $g_{NL}$ required to generate this asymmetry. We note that there are models with a large $g_{NL}$ and small $f_{NL}$, for example, \cite{Elliston:2012wm} and \cite{Enqvist:2014nsa}.
A requirement on the model is that the isocurvature field contributes a small amount towards the power spectrum of the curvature perturbation, which could be considered 
a fine tuning. We also note that the large minimal value of $g_{NL}$ required implies our observable patch of the universe has a significantly smaller value of $\fnl$ than our neighbours. The observed asymmetry is scale dependent, with a smaller asymmetry on small scales, which this model does not account for.

If the observed asymmetry is due to the higher order term considered in this work, then this will put strong bounds on $f_{NL}$ and $g_{NL}$. Measurements of $f_{NL}$ and $g_{NL}$ outside of our allowed region would falsify models which use this cubic term to generate the asymmetry.

The cubic term has a different $\vect{\hat{n}}$-dependence compared to the first order term. For this paper we assumed $A_3\gae 0.07$, but we would like to see a fit to the data with the $m=1,2,3$ terms, in order to properly constrain the parameters $A_1, A_2$ and $A_3$. 

This study has shown that a higher order term can generate the required asymmetry, relaxing the constraint on $\fnl$ compared to that generated only by the first order. Perhaps the other cubic order term in \eqref{pzeta3}, $N_{,\chi\chi\chi\chi}N_{,\chi}$, may also contribute -- although the bound on the non-linear parameter, $h_{NL}$ \cite{2011JCAP...07..033S}, associated to this term is considerably weaker than that on $g_{NL}$, and so this term is not as easily falsifiable. Indeed, since the third order term can have a large contribution, other higher order terms (and combinations of them) may also be significant. Our work prompts investigation of the case where $\delta N$ can't be Taylor expanded.

\section{Acknowledgements}

We would like to thank Takeshi Kobayashi and Christian T. Byrnes for many useful comments and discussions on a draft of this work. ZK is supported by an STFC studentship. DJM is supported by a Royal Society University Research Fellowship. ST is supported by STFC consolidated grants ST/J000469/1 and ST/L000415/1.

\bibliography{asymmbib23_04.bib}

\end{document}